\documentstyle[12pt]{article}
\newlength{\myleftmargin}
\newlength{\paperwidth}
\newcommand{\an}{ans{\" a}tze}
\newcommand{\aaa}{2^{-\frac{1}{4}}}
\newcommand{\aaaa}{2^{-\frac{3}{4}}}
\newcommand{\ccc}{\circ}
\begin{document}
\thispagestyle{empty}

\begin{flushright}
{\tt DPNU-96-42 \\
August 1996}
\end{flushright}

\vspace{5em}

\begin{center}
{\Large{\bf The Unitarity Triangle on the First Quadrant}}
\end{center}
\begin{center}
{\Large{\bf and the Quark Mass Matrices}}
\end{center}
\begin{center}
{\Large{\bf in the Nearest-Neighbor Interaction Basis}}
\end {center}

\vspace{2em}

\begin{center}
{\sc Toshiaki~~Ito}\\
\sl{Department of Physics, Nagoya University,} {\it Nagoya 464-01, Japan}
\end{center}

\vspace{4em}

{\large{\bf Abstract :}} The unitarity triangle
on the first quadrant in the $\rho -\eta$ plane is discussed
in the framework of the quark mass matrices
in the NNI basis.
If the quark mass matrices of the up-type is the Fritzsch
one and the down-type is
the one proposed by  Branco et al.,
respectively, one gets the unitarity triangle
with the vertex on the first quadrant.
This simple model may be a candidate
of the quark mass matrix {\an},
if the allowed region of the vertex of the unitarity
triangle is restricted in the first quadrant by future experiments.

\clearpage

One of the most important problems of flavor physics is to
understand flavor mixing and fermion masses, which
are free parameters in the standard model.
The observed values of those mixing and masses provide us with
clues of the origin of the fermion mass matrices.
One of the most stringent tests of the quark mass matrices
is an examination of the so called unitarity triangle of the
Kobayashi-Maskawa(KM) matrix[1].
Then one needs the experimental information of the six quark masses
to estimate the KM matrix elements by quark mass matrix models.
The discovery of top quark[2] can lead to
the precise study of the quark mass matrices. 
Thus, we are now in the epoch of examining the quark mass matrices
in terms of KM matrix elements.

As presented by Branco, Lavoura and Mota, both up and down
quark mass matrices could always be transformed
to the non-hermitian matrices
in the nearest-neighbor interaction (NNI) basis
by a weak-basis transformation for the three and four generation cases[3,4].
In this basis,  the KM matrix elements are expressed generally
in terms of mass matrix parameters due to eight texture zeros.
In particular, phases of the mass matrices can be easily isolated.
The famous Fritzsch {\an}[5] is the special one of the NNI basis.
This {\an} is viable for the $|V_{us}|$ element as follows
\begin{equation}
V_{us} \simeq -\sqrt{\frac{m_d}{m_s}}e^{ip}+
\sqrt{\frac{m_u}{m_c}}e^{iq}, \label{eqn:fvus}
\end{equation}
where $p$ and $q$ are phase parameters.

In the ref[6],
we have examined the unitarity triangle from the quark mass matrices
with the generation hierarchy in the NNI basis.
It is emphasized that the position of vertex of the unitarity triangle
is on the second quadrant of the $\rho -\eta$
plane[7] as far as Eq.(\ref{eqn:fvus}) holds.
However, the experimentally allowed region of
the triangle vertex also exists on the first quadrant
as well as the second one.
So, in this paper,
we consider the quark mass matrices that the vertex
position of the unitarity triangle stays on the first quadrant.
In the ref.[6], it is shown that
there is a possibility that the vertex of the
unitarity triangle moves into the first quadrant
if the large discrepancy from Fritzsch {\an} will be obtained
in future experiments.
We propose a simple example of the quark
mass matrices which implies the unitarity triangle with the
first quadrant vertex.

Let us begin with considering
two typical  matrices for the quark mass matrices,
\begin{equation}
M_1=
\left(
\begin{array}{ccc}
0 & A & 0 \\
A^* & 0 & B \\
0 & B^* & C
\end{array}
\right)
~~~,~~
M_2=
\left(
\begin{array}{ccc}
0 & D & 0 \\
D^* & 0 & E \\
0 & F & F
\end{array}
\right)
~~~,
\end{equation}
where $A$, $B$, $D$, $E$ and $F$ are complex numbers, 
while $C$ is a real numbers.
Then, there are following four possible  cases in principle
by combining two type  matrices,
\begin{center}
\begin{tabular}{lll}
case I & $M_u=M_1$ & $M_d=M_1$, \\
case II & $M_u=M_1$ & $M_d=M_2$, \\
case III & $M_u=M_2$ & $M_d=M_1$, \\
case IV & $M_u=M_2$ & $M_d=M_2$.
\end{tabular}
\end{center}

The case I is the well known Fritzsch {\an}[5].
Although this {\an} is successful for the
$V_{us}$ element as shown in Eq.(\ref{eqn:fvus}),
it fails for $V_{cb}$ as far as $m_t\geq 100$GeV.
On the other hand, the case IV which is the {\an} proposed by Branco et al.[8]
is successful not only for the $V_{us}$ element but also
for the $V_{cb}$ element.
Although this {\an} overcomes the fault of the Fritzsch {\an},
it cannot reproduce the observed ratio of $|V_{ub}|/|V_{cb}|=0.08\pm 0.02$[9].
Actually, the case IV predicts the larger value than 0.12 for this ratio.
Then we examine other two cases.
First one is  the case II;
\begin{equation}
M_u=M_1~~~,~~~M_d=M_2. \label{eqn:caseone}
\end{equation}
The matrices $U_u$ and $U_d$ are defined
as the unitarity matrices which diagonalize
the hermitian matrices $H_u=M_u$ and $H_d=M_dM_d^{\dagger}$,
respectively,
\begin{equation}
U^{\dagger}_uH_uU_u=D_u~~, \qquad
U^{\dagger}_dH_dU_d=D_d~,
\end{equation}
where $D_u={\rm diag.}(m_u,m_c,m_t)$ and
$D_d={\rm diag.}(m_d^2,m_s^2,m_b^2)$.
In the NNI basis, we can extract phases from each quark mass matrix
by the use of the diagonal phase matrices.
Since phases of the mass matrices can be isolated, we are able to write
\begin{equation}
U_u=\phi_uO_u~~, \qquad U_d=\phi_dO_d \label{eqn:orth}
\end{equation}
where $\phi_u={\rm diag}.(e^{ip_u},e^{iq_u},1)$,
$\phi_d={\rm diag}.(e^{ip_d},e^{iq_d},1)$ and $O_u$, $O_d$ are
orthogonal matrices.
We define the phase matrix,
${\sl\Phi}=\phi_u^*\phi_d={\rm diag.}(e^{ip},e^{iq},1)$
with $p=p_d-p_u$ and $q=q_d-q_u$.
Then the KM matrix is given by,
\begin{equation} 
V_{KM}=U_u^{\dag}U_d=O_u^T{\sl\Phi}O_d \ .
\end{equation}
In the case of Eq.(\ref{eqn:caseone}),
we obtain the KM matrix elements approximately,
\begin{eqnarray}
V_{ud} &\simeq& \frac{1}{N_{ud}}\left( e^{ip}+\aaa \sqrt{\frac{m_um_d}{m_cm_s}}
e^{iq}+\aaaa \sqrt{\frac{m_um_dm_s}{m_tm_b^2}}\right), \label{eqn:vud} \\
V_{us} &\simeq& \frac{1}{N_{us}}\left( -\aaa\sqrt{\frac{m_d}{m_s}}e^{ip}
+\sqrt{\frac{m_u}{m_c}}e^{iq}+\sqrt{\frac{m_u}{m_t}}\frac{m_s}{m_b}
\right), \label{eqn:vus} \\
V_{ub} &\simeq& \frac{1}{N_{ub}}\left( \aaa\sqrt{\frac{m_dm_s}{m_b^2}}e^{ip}
+\sqrt{\frac{m_u}{m_c}}
\frac{m_s}{m_b}e^{iq}-\sqrt{\frac{m_u}{m_t}}\right), \label{eqn:vub} \\
V_{cd} &\simeq& \frac{1}{N_{cd}}\left( \sqrt{\frac{m_u}{m_c}}e^{ip}
+\aaaa\sqrt{\frac{m_d}{m_s}}e^{iq}
+\aaa\sqrt{\frac{m_cm_dm_s}{m_tm_b^2}}\right), \label{eqn:vcd} \\
V_{cs} &\simeq& \frac{1}{N_{cs}}\left(-\aaa\sqrt{\frac{m_um_d}{m_cm_s}}e^{ip}
+e^{iq}+\sqrt{\frac{m_c}{m_t}}\frac{m_s}{m_b}\right), \label{eqn:vcs} \\
V_{cb} &\simeq& \frac{1}{N_{cb}}\left( \aaa\sqrt{\frac{m_um_dm_s}{m_cm_b^2}}
e^{ip}+\frac{m_s}{m_b}e^{iq}-\sqrt{\frac{m_c}{m_t}}\right), \label{eqn:vcb} \\
V_{td} &\simeq& \frac{1}{N_{td}}\left( \frac{m_c}{m_t}\sqrt{\frac{m_u}{m_t}}
e^{ip}+\aaa\sqrt{\frac{m_dm_c}{m_sm_t}}e^{iq}-\aaaa\sqrt{\frac{m_dm_s}{m_b^2}}
\right), \label{eqn:vtd} \\
V_{ts} &\simeq& \frac{1}{N_{ts}}\left( -\aaa \frac{m_c}{m_t}
\sqrt{\frac{m_um_d}{m_tm_s}}e^{ip}+\sqrt{\frac{m_c}{m_t}}e^{iq}
-\frac{m_s}{m_b}\right), \label{eqn:vts} \\
V_{tb} &\simeq& \frac{1}{N_{tb}}\left( \aaa \frac{m_c}{m_t}
\sqrt{\frac{m_um_dm_s}{m_tm_b^2}}e^{ip}+\sqrt{\frac{m_c}{m_t}}\frac{m_s}{m_b}
e^{iq}+1 \right), \label{eqn:vtb}
\end{eqnarray}
where $1/N_{ij}$s are normalization factors.

In order to estimate the absolute values of
KM matrix elements $|V_{ij}|$, we must know the values
of the masses of six quarks on the same energy scale.
Following the study by Koide[10],
we obtain the values of quark masses
at 1GeV
by using the 2-loop renormalization group equations,
\begin{eqnarray*}
m_u &=& 0.0056\pm 0.011~~,~m_d=0.0099\pm 0.0011~~,~m_s=0.199\pm 0.033~~, \\
m_c &=& 1.316\pm 0.024~~,~m_b=5.934\pm 0.101~~,
~m_t=349.5\pm 27.9~({\rm GeV})~~,
\end{eqnarray*}
where $\Lambda^{(5)}_{\overline{MS}}=0.195$GeV.
Hereafter, we use the central values for the numerical estimation.
If we put the phase parameters on $p=-73^{\ccc}$ and $q=47^{\ccc}$,
we obtain the absolute values of the KM matrix elements;
\begin{equation}
|V_{KM}| = 
\left(
\begin{array}{ccc}
0.97556 & 0.21971 & 0.00455 \\
0.21957 & 0.97452 & 0.04580 \\
0.00917 & 0.04510 & 0.99894
\end{array}
\right). \label{eqn:eskm}
\end{equation}
All of these nine values are consistent with the experimental data[9] as;
\begin{equation}
|V_{KM}| =
\left(
\begin{array}{ccc}
0.9745~{\rm to}~0.9757 & 0.219~{\rm to}~0.224 & 0.002~{\rm to}~0.005 \\
0.218~{\rm to}~0.224 & 0.9736~{\rm to}~0.9750 & 0.036~{\rm to}~0.046 \\
0.004~{\rm to}~0.014 & 0.034~{\rm to}~0.046 & 0.9989~{\rm to}~0,9993
\end{array}
\right).
\end{equation} 
The unitarity triangle obtained from the KM matrix elements
of the Eq.(\ref{eqn:eskm}) is shown in Fig.1.
Here, in order to describe the experimentally allowed region,
we used the recent JLQCD result[11] of the Lattice Calculation
for the theoretical parameters $\hat B_K$ and $f_{B_d}$
as follows:
\begin{equation}
\hat B_K = 0.76\pm 0.04~, \qquad\quad f_{B_d} = 0.19\pm 0.01 {\rm GeV}~.
\end{equation}
The position of the vertex point of this unitarity triangle
is on the first quadrant of the $\rho -\eta$ plane.
So the case II is an {\an} which put the unitarity triangle
on the first quadrant.

Next, we consider the case III;
\begin{equation}
M_u=M_2~~~,~~~M_d=M_1. \label{eqn:casetwo}
\end{equation}
The KM matrix elements are obtained analogously.
For example, $V_{cb}$ is given as,
\begin{equation}
V_{cb} = -2^{-\frac{1}{4}}\frac{m_s}{m_b}\sqrt{\frac{m_um_d}{m_cm_b}}
e^{ip}+\sqrt{\frac{m_s}{m_b}}e^{iq}-\frac{m_c}{m_t}.
\end{equation}
Unfortunately, we cannot reproduce the experimental value of $|V_{cb}|$
even if any values of the phase parameters are taken.

In this paper, we examine the unitarity triangle on the first quadrant
of the $\rho -\eta$ plane by using the quark mass matrices in 
the NNI basis.
Assuming $(M_u)_{12}=(M_u)^*_{21}$ and $(M_d)_{12}=(M_d)^*_{21}$,
four cases(I$\sim$IV) are considered.
There is only one case, {\it i.e.} case II, which can reproduce
all absolute values of the KM matrix elements.
The quark mass matrix combination of the case II implies the unitarity
triangle with the vertex on the first quadrant.
If the $B$-factory experiments at KEK and SLAC will restrict
the experimentally allowed region of the unitarity triangle
in the first quadrant of the $\rho -\eta$ plane,
our proposed simple model can be a candidate
for the quark mass matrices.

\vspace{3em}

{\bf Acknowledgments}

I would like to thank Professor M. Tanimoto and Professor M. Matsuda
for helpful comments and useful advice.

\vspace{3em}

{\bf Figure Captions}

{\bf Fig.1} : The unitarity triangle when the phase parameters are
put on $p=-73^{\ccc}$ and $q=47^{\ccc}$ in the case II.

\end{document}